\documentclass[rnote,oldversion]{aa}  

\usepackage{graphicx}
\usepackage{txfonts}

\begin{document}

   \title{Seismic evolution of low/intermediate mass PMS stars}

   \subtitle{}

   \author{F. J. G. Pinheiro
          \inst{1}
          }

   \offprints{F. J. G. Pinheiro (fjgp@astro.up.pt)}

   \institute{Centro de Astrof\'{\i}sica da Universidade do Porto 
             , Rua da Estrelas , 4150-762 Porto, Portugal}
   \date{ }

  \abstract{
 This article presents a study of the evolution of the internal 
structure and seismic properties expected for low/intermediate 
mass \hbox{Pre-Main} Sequence (PMS) stars. 
 Seismic and non-seismic properties of PMS stars were analysed.
 This was done using
 0.8 to 4.4M$_\odot$ stellar models at stages ranging from the 
end of the Hayashi track up to the Zero-Age Main-Sequence (ZAMS). 
 This research concludes that, for intermediate-mass stars 
(M$>$1.3M$_\odot$), diagrams comparing the effective temperature 
($T_{eff}$) against the small separation can provide an alternative to 
Christensen-Dalsgaard (C-D) diagrams.
 The impact of the metal abundance of intermediate mass stars 
(2.5-4.4M$_\odot$) has over their seismic properties is also
evaluated. 
}
   \maketitle
  

\section{Introduction}

 The study of the Sun's seismic characteristics allows 
a better understanding of its internal structure 
(e.g. Antia \cite{antia05}).
 Yet, for other stars it is only possible to observe 
pulsations associated with low degree spherical harmonics,   
constraining the results that one can obtain (Bedding \& 
Kjeldsen \cite{bedding03}). 
 Still, asteroseismic techniques can be used to infer the 
internal structure of solar-type stars (e.g. Monteiro et 
al., \cite{monteiro02})
 
 The seismic study of Pre-Main Sequence (PMS) stars (e.g. 
Breger \cite{breger72}, Marconi \& Palla \cite{marconi98}, 
Ripepi \& Marconi \cite{ripepi04}) has been mainly focused 
on objects located inside the PMS Instability Strip (IS).  
 However, in terms of the location and size of the outer 
convective regions, low mass PMS stars, located near the 
Zero Age Main Sequence (ZAMS), resemble solar-type stars. 
 Therefore they may be expected to oscillate. 

 Not much work has been done regarding the study of PMS 
solar-type pulsations (Samadi et al. \cite{samadi05}, 
Pinheiro et al. \cite{pinheiro06}). 
 This research goes beyond the study of the seismic properties 
of young stars initiated by Pinheiro et al. (\cite{pinheiro06})
 by extending  Monteiro et al.'s (\cite{monteiro02}) analysis 
to PMS stars. 
 By taking into account the evolution of the seismic properties 
of young stars we evaluate the possibility of testing PMS 
evolutionary models through the use of solar-type pulsations.

\section{Solar-type pulsations}

 The Sun, amongst other solar-type stars, displays a type of 
oscillation in which pressure acts as a restoring force.
 These \hbox{p-mode} pulsations, known as solar-type oscillations,
are stochastically driven by the outer convective layers. 
 The power spectrum of solar-type pulsators presents two 
ty\hbox{pical} frequency separations, known as the large 
($\Delta\nu_{n,l}$) and small ($\delta\nu_{n,l}$) separation. 
  The first one corresponds to the difference between 
frequencies associated with oscillations of the same degree ($l$) 
and consecutive overtone ($n$), while the latter corresponds 
to the difference between frequencies $\nu_{n-1,l+2}$ 
and $\nu_{n,l}$. 
 Tassoul (\cite{tassoul80}) showed that in the asymptotic 
regime ($n\gg l$) these two separations can be written as:

\begin{equation}
  \label{eq:eq1}
  \Delta\nu_{n,l} = \nu_{n+1,l} - \nu_{n,l} \propto \Delta\nu = 
     \bigg( 2 \int_{0}^{R} \frac{dr}{C_s} \bigg)^{-1} =
  \bigg( 2 \int_{0}^{R} \frac{dr}{\sqrt{\Gamma_1 p/\rho}} \bigg)^{-1}
\end{equation}

\begin{equation}
  \label{eq:eq2}
   \delta\nu_{n,l} = \nu_{n-1,l+2} - \nu_{n,l} \propto 
 \frac{\Delta\nu}{\nu_{n,l}} \times  r_{i} ~ ~ ~ \mathrm{and} 
~ ~ ~ r_{i} = \int_{0}^{R} \frac{\partial C_s}{\partial r} \frac{dr}{r} ~ ~, 
\end{equation}

\noindent where $\Gamma_1$, p and $\rho$ are the 
adiabatic exponent, pressure and density inside the star. 
 Therefore $\Delta\nu$ is a measurement of 
the inverse of the time that acoustic waves, with a 
velo\hbox{city} $Cs$, take to travel from the centre to 
the stellar surface. 
 According to Kjeldsen \& Bedding (\cite{kjeldsen95}), for a 
fully ionised star composed of an ideal gas with an ``average'' 
temperature $<$T$>\propto$M$^2$/R$^4$, one gets: 

\begin{equation}
  \label{eq:eq3}
     \Delta\nu \propto \sqrt{\mathrm{M/R}^3} \propto \sqrt{<\rho>} 
= \sqrt{ \frac{1}{4/3  ~ \pi R^{3}} ~ 
{\textstyle \int }_{0}^{R} 4\pi ~ \rho(r) ~ r^{2} dr    }  ~ ,
\end{equation}

\noindent i.e. $\Delta\nu$ depends on a global stellar 
parameter  ($<$$\rho$$>$).
 The sound speed gradient integral $r_{i}$ (defined on Eq. 
\ref{eq:eq2}) is related to the \hbox{astero}seismic ratio  
$r_{n,l}$$=$$\delta\nu_{n,l}/\Delta\nu_{n,l+1}$ 
(Roxburgh \& Vorontsov \cite{roxburgh03}).  
 Equation 2 hints that  $r_i$ and $\delta\nu_{n,l}$ are sensitive 
to the sound speed gradient ($\partial C_s/\partial r$) near the 
stellar centre.  Thus $\delta\nu_{n,l}$ is sensitive 
to the stellar structure and evolutionary status.
 Diagrams which compare these two frequency separations can be 
used to study solar-type stars 
(Christensen-Dalsgaard \cite{dalsgaard84}). 
 These are known as Christensen-Dalsgaard (C-D) diagrams.

\section{Evolution of the seismic properties}

\begin{figure}[htb]
  \begin{center}
   \includegraphics[angle=90,width=12.6cm,height=10.cm]{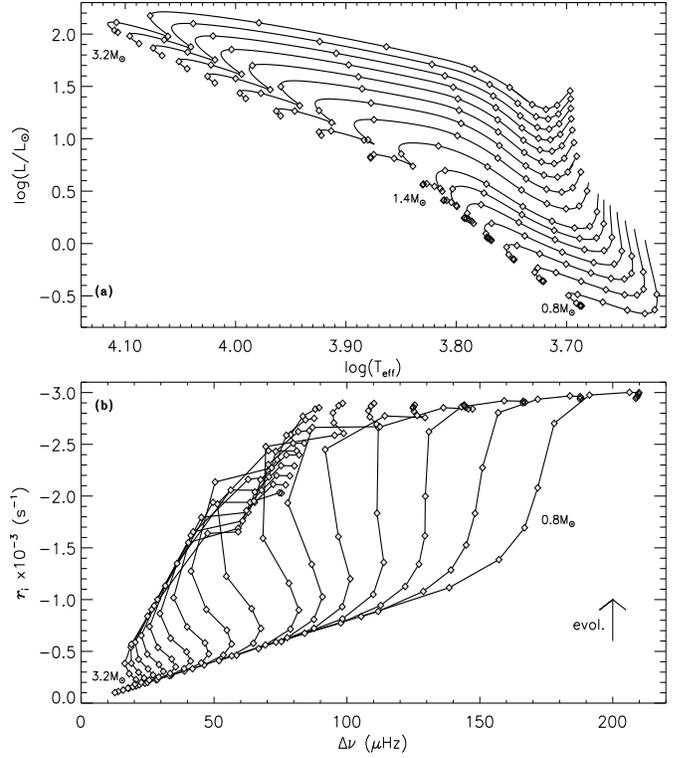}
    \caption{
{\bf{a)}} PMS evolutionary tracks for 0.8 to 3.2M$_\odot$ stars. 
 The diamonds correspond to the models described in Sect. 3. 
{\bf{b)}}  C-D diagram produced using the same models. 
 The arrow indicates the direction in which evolution takes place. }
    \label{fig:1}
  \end{center}
\end{figure}

 Stellar models are required to analyse the evolution of 
the internal structure and seismic properties of PMS stars. 
 PMS \hbox{models} produced in preparation for the COROT mission 
(available at \textit{http://www.astro.up.pt/corot/models}) 
were used in this study.
 These models were produced using the CESAM stellar 
evolutionary code (Morel \cite{morel97}), 
 applying the same procedures and assuming the same 
physical ingredients used 
in the modelling of EK Cep (Marques et al. \cite{marques04}).
 These  Z=0.02 models have masses between 0.8 and 3.2M$_\odot$. 
 For each mass, 13 models were available.  
 These are equally spaced in time, \hbox{ranging} from 
the end of the Hayashi track up to the ZAMS.
 Figure \ref{fig:1}.a shows each model's position in 
the HR diagram. 

 The model's seismic parameters ($\Delta\nu$ \& $r_{i}$) were 
estimated using the integrals shown in Eqs. 1 and 2. 
 The \hbox{C-D} diagram displayed in Fig. \ref{fig:1}.b 
was produced following Roxburgh \& Vorontsov's approach 
(\cite{roxburgh03}), i.e. $r_{i}$ was used instead 
of $\delta\nu_{n,l}$. 
 The overlap of different models on the C-D diagram implies 
that this diagram is only useful to analyse stars with less 
than 1.3M$_\odot$.

\subsection{Sound speed gradient integral vs. effective temperature} 

\begin{figure}[hbt]
  \begin{center}
   \includegraphics[angle=90,width=12.cm,height=9.35cm]{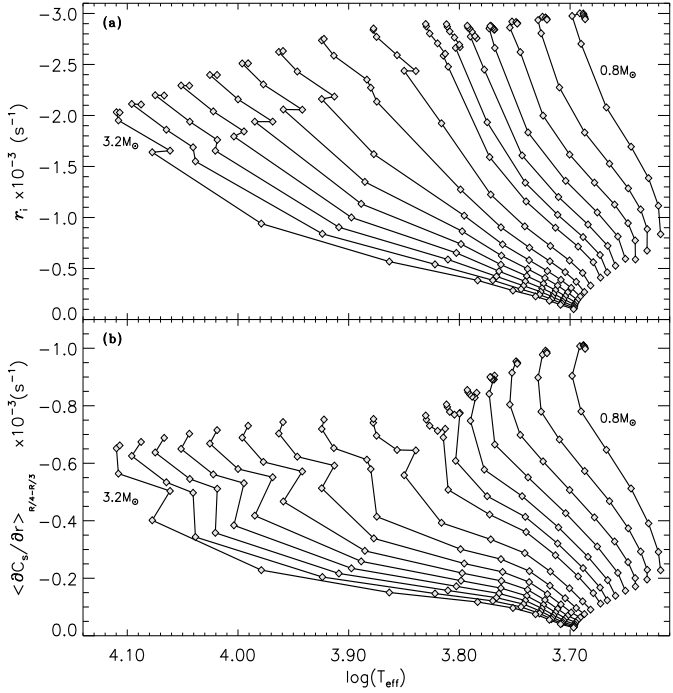}
    \caption{ 
{\bf{a)}} Evolution of the sound speed gradient integral 
and {\bf{b)}} the \hbox{average} sound speed gradient 
$<\partial{C_s}/\partial{r}>_{R/4-R/3}$ 
for the models displayed at Fig. 1 (diamonds).
}
    \label{fig:figr2}
  \end{center}
\end{figure}

 By comparing, for each model, the effective temperature 
(T$\mathrm{_{eff}}$)
against its $r_i$ integral (Fig. \ref{fig:figr3}.a) one 
breaks the small and large separation degeneracy displayed 
by some of the models (Fig. \ref{fig:figr2}.b). 
 This means that unlike C-D diagrams, T$_{eff}$ vs. $r_i$ 
diagrams can be used to study objects more massive than 
1.3M$_\odot$.
 Yet Monte Carlo simulations point out that, for low mass stars 
(M$<$$1.3$M$_\odot$), $T_{eff}$ vs. $r_i$ diagrams can 
only achieve the precision of C-D diagrams if  $T_{eff}$ is 
accurately known. 
 For instance, a 3.5$\%$ accuracy in the mass determination 
is achieved either by knowing $r_i$ and $\Delta\nu$ with a 
5$\%$ precision or by knowing $r_i$ (with that accuracy) 
and T$_{eff}$ with a 50K uncertainty.

 Figure \ref{fig:figr2} shows that the $r_i$ integral 
evolves in a similar way as the average sound speed 
gradient computed between 0.25 and 0.33 stellar radii 
$\left( <\partial{C_s}/\partial{r}>_{R/4-R/3} = 
\frac{1}{R/3-R/4} \int_{R/4}^{R/3} 
            \frac{\partial C_{s}}{\partial r} dr \right)$.
 This hints at $r_i$'s dependence on the inner sound 
speed gradient.

\subsection{The large separation and the mean stellar density} 

 From Eq. 3 we get that $\Delta\nu$$\propto$$<$$\rho$$>^{1/2}$. 
 By relying on the definition of 
$\Delta\nu$ and $<$$\rho$$>^{1/2}$(Eqs. 1 \& 2) we can 
write the $\Delta\nu$$/$$<$$\rho$$>^{1/2}$ ratio as:

\begin{displaymath}
  \Delta\nu/<\rho>^{1/2} = 
 \frac{1}{\frac{2}{C_{S}(0)} \int_{0}^{R} \frac{C_{S}(0)}{C_{S}(r)} dr  }
 \times  
 \left( \frac{4/3 ~ ~ \pi R^3}{\rho(0)\int_{0}^{R} 4\pi 
                 \frac{\rho(r)}{\rho(0)} r^2 dr  } \right)^{0.5}
 \stackrel{ x = \frac{r}{R}  }{=} 
\end{displaymath}
\begin{displaymath}
 ~ ~  \stackrel{ x = \frac{r}{R}  }{=} 
   \left(\frac{ C_{S}(0)}{2R \sqrt{3 \rho(0) } }\right) \times
   \left(\int_{0}^{1}\frac{C_{S}(0)}{C_{S}(x)} dx \right)^{-1}\times
   \left(\int_{0}^{1} \frac{\rho(x)}{\rho(0)} x^{2} dx  \right)^{-0.5}=
\end{displaymath}

\begin{equation}
  \label{eq:eq4}
 ~ ~ ~ \, =  K_{0} \times {I_{C}}^{-1} \times {I_{\rho}}^{-0.5}  ~ , 
\end{equation}

\noindent where  $K_{0}$ is an estimate of 
$\Delta\nu$/$<\rho>^{1/2}$ obtained using the central 
density ($\rho(0)$) and sound speed ($C_{S}(0)$). 
 The normalised sound speed ($I_{C}$) and density 
($I_{\rho}$) integrals are independent of the model's 
radius, density and sound speed.

\begin{figure}[hbt]
  \begin{center}
   \includegraphics[angle=90,width=24.cm,height=18.7cm]{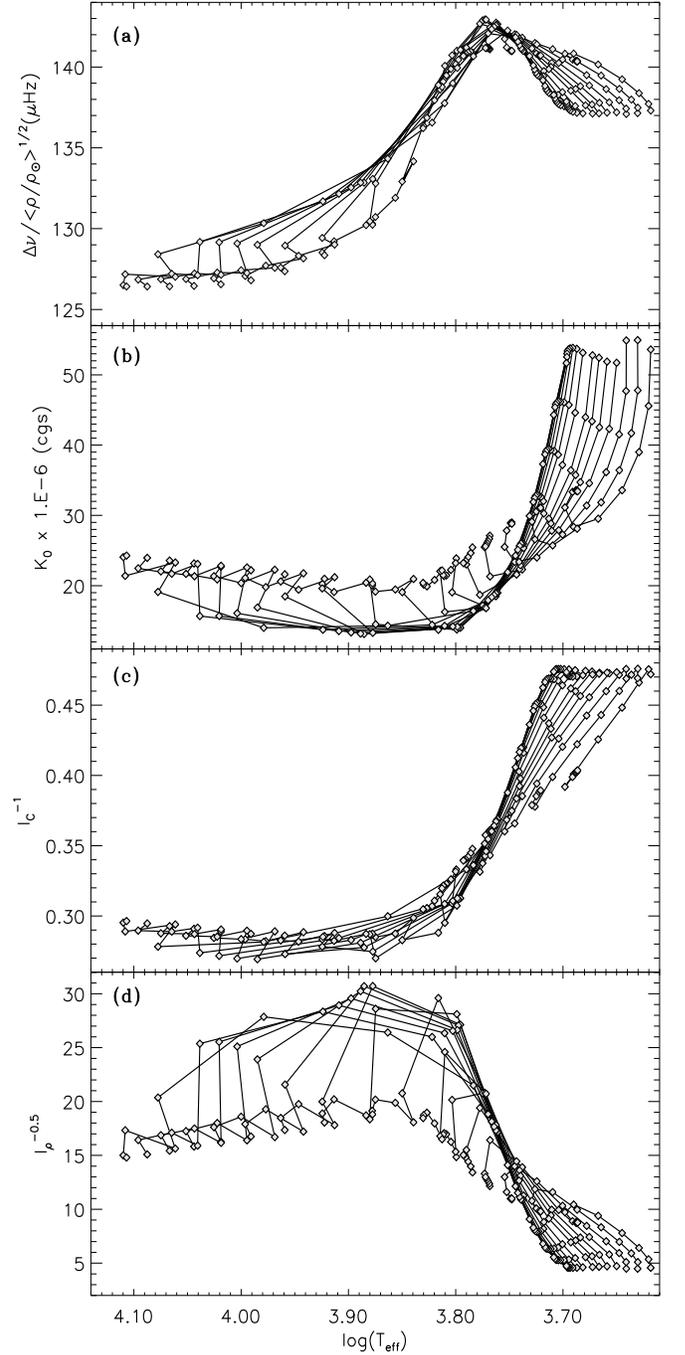}
   \caption{ 
{\bf{a)}} Normalization of the large separation ($\Delta\nu$) by
the square root of the mean stellar density $<$$\rho$$>^{1/2}$,  
{\bf{b)}} $\Delta\nu$/$<$$\rho$$>^{1/2}$ ratio estimated using 
the core density and sound speed, 
{\bf{c)}} inverse of the profile integral $I_C$ and,
{\bf{d)}} inverse of the square root of the profile integral 
$I_\rho$ for the models described in Sect. 3 (diamonds).
}
    \label{fig:figr3}
  \end{center}
\end{figure}

 Figure \ref{fig:figr3} hints 
that changes in the density and sound speed profiles 
are partially responsible for the observed variations in the 
$\Delta\nu$/$<$$\rho$$>^{1/2}$ ratio. 
 Such profile changes are due to modifications in the 
relationship between pressure and density resul\hbox{ting} 
from changes in the location/size of convective/ionisation 
regions that occur during the contraction towards the ZAMS.

 This figure also shows that $\Delta\nu$/$<$$\rho$$>^{1/2}$, 
$K_{0}$, $I_{C}$ and $I_{\rho}$ are, to some extent, 
correlated with the model's effective temperature. 
  Indeed, the dispersion of the model's 
$\Delta\nu$/$<$$\rho$$>^{1/2}$ ratio in Fig. \ref{fig:figr3}.a 
is, at any given temperature, around 2 to 5\% 
 Thus knowing a star's effective temperature allows one 
to predict its $\Delta\nu$/$<$$\rho$$>^{1/2}$ ratio.

\section{The contribution of metallicity}           

 Intermediate mass PMS stars are known for their 
$\delta$-Scuti type pulsations.  
 Samadi et al. (\cite{samadi05}) predicted the 
amplitudes of solar-type pulsations for  
low mass PMS stars located to the right of the IS.
 The size (in solar radius) of the outer convective 
region of Samadi's PMS model \#9 is smaller than that 
of the log(T$\mathrm{_{eff}}$)$\leq$3.85 models used 
in Sect. 3.
 Therefore, one may be tempted to extrapolate Samadi's 
result for \hbox{log(T$\mathrm{_{eff}}$)$\leq$3.85 stars}. 
 On the other hand, intermediate mass stars located near 
the ZAMS may not display solar-type pulsations since their 
outer convective layers are smaller (about one order of 
magnitude) than the one of Samadi's PMS model \#9. 
 The convective cores of A type stars ($\sim$2M$_\odot$) 
can drive g modes (Browning et al. \cite{browning04} 
\& Antonello et al. \cite{antonello06}). 
 However, currently it is unknown whether core convection 
can drive solar-type pulsations like outer convection does. 
 Nonetheless, such conjectures require an extrapolation 
of the study of solar-type pulsations to more 
massive stars.

\begin{figure}[hbt]
  \begin{center}
   \includegraphics[angle=90,width=12.cm,height=9.35cm]{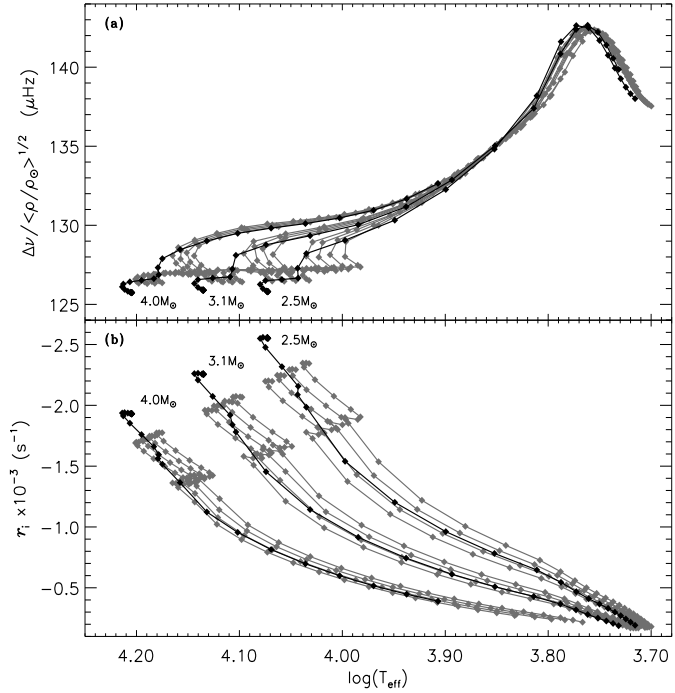}
   \caption{ 
{\bf{a)}} Evolution of the $\Delta\nu$/$<\rho>^{1/2}$ ratio and  
{\bf{b)}} $r_{i}$ integral for diferent stellar models. 
The black diamonds correspond to 2.5M$_{\odot}$, 3.1M$_{\odot}$ 
and 4.0M$_{\odot}$ Z=0.01 models. 
The grey diamonds  correspond to 2.5-2.8M$_{\odot}$, 
3.1-3.4M$_{\odot}$ and 4.0-4.3M$_{\odot}$ Z=0.02 models.
}
    \label{fig:figr4}
  \end{center}
\end{figure}

 In order to explore the metalicity effect on the 
seismic properties of intermediate mass stars, we 
selected Z=0.01 and Z=0.02 models produced by J. 
Marques to study the PMS \hbox{$\delta$-Scuti} star 
VV Ser (Ripepi et al. \cite{ripepi07}).
 These models have masses ranging between 2.5 and 
4.4M$_\odot$ covering, for each mass, 30 evolutionary 
stages equally spaced in time. 
 Figure \ref{fig:figr4} shows that the seismic properties 
of Z=0.02 PMS stars evolve in a similar fashion to those of 
Z=0.01 stars. 
 This happens because the internal  structure of both 
types of stars evolves in the same way. 
 However quantitatively there are some \hbox{diffe}rences.
 Near the ZAMS, metal-poor stars are denser and have larger 
sound speed gradients than their Z=0.02 counterparts. 
 Therefore their $r_i$ integral is larger.

\section{Asteroseismic test of PMS models}

\begin{figure}[bht]
  \begin{center}
   \includegraphics[angle=90,width=12.cm,height=9.35cm]{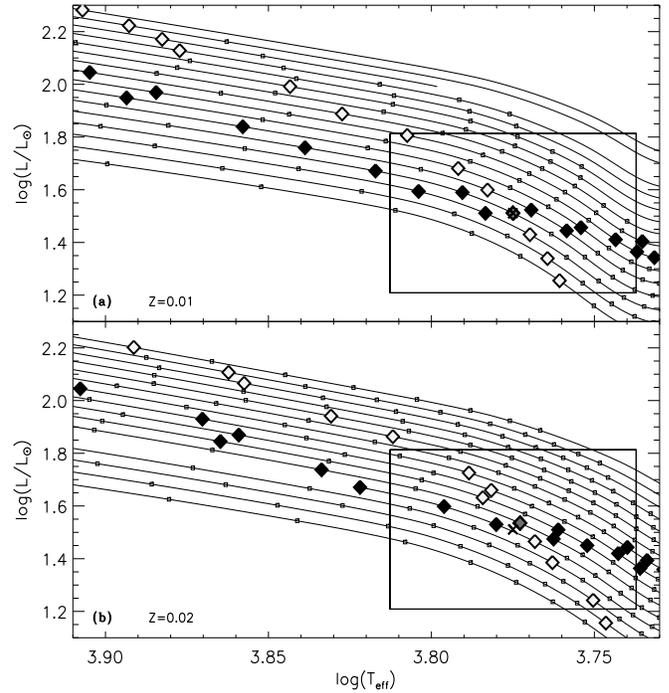}
    \caption{
{\bf{a)}} Z=0.01 and, {\bf{b)}} Z=0.02 stellar models 
that reproduce, with 5\% accuracy, the large separation 
$\Delta\nu$ (black diamonds) and the $r_i$ integral (white 
diamonds) of a 2.8M$_\odot$, Z=0.01 test star whose 
position in the HR diagram is given by the black cross 
(log(T$_{eff}$)=3.77 \& log(L/L$_{\odot}$)=1.51). 
 The grey diamonds correspond to the models that 
reproduce both seismic properties with a 5\% accuracy, 
while the small squares represent models that do not 
reproduce them.
 The boxes represent a 1\%$\times$20\%  uncertainty 
region (in log(T$_{eff}$) and log(L/L$_\odot$)) 
around the black crosses. 
 The lines are 2.5 to 4.0M$_\odot$ PMS evolutionary tracks.}
    \label{figr5}
  \end{center}
\end{figure}

 The seismic properties of a given star can be used 
to test stellar models that reproduce, within given 
uncertainties, its position in the HR diagram.
 Figure \ref{figr5} shows this. 
 In this \hbox{exercise} the target is a 2.8M$_\odot$, 
Z=0.01 star with log(L/L$_{\odot}$)=1.51 and 
log(T$_{eff}$)=3.77.
 The models tested here are the same ones used in 
Sect. 4.  
 In this case a 20\% uncertainty is assumed in the 
target's log(L/L$_{\odot}$), a 1\% uncertainty in 
log(T$_{eff}$) and a 5\% accuracy in its 
seismic parameters $\Delta\nu$ and $r_i$.
 Figure \ref{figr5} hints that, as expected from Eq. 3, 
$\Delta\nu$ puts \hbox{cons}traints on the model's density. 
 In the same manner we can see how  $r_i$ puts 
\hbox{cons}traints on the stellar evolutionary status. 
 Conversely, models that reproduce the seismic properties 
of a given star can constrain its global stellar parameters 
(e.g. Pinheiro et al. \cite{pinheiro03}).

 According to Baglin et al. (\cite{baglin01}), in a 150 day 
run COROT can achieve a 0.1$\mu$Hz accuracy in individual 
frequency determination.
 On the other hand, $\Delta\nu_{n,l}$$\approx$40$\mu$Hz and 
$\delta\nu_{n,l}$$\approx$6$\mu$Hz are typical 
frequency separations of intermediate mass PMS stars 
(Pinheiro et al. \cite{pinheiro06}). 
 This means that COROT can determine $\Delta\nu_{n,l}$ and 
$\delta\nu_{n,l}$ with accuracies up to 0.5\% and
4\%, respectively.
 Table \ref{tab:1} shows that some of the models used above 
have different seismic properties, despite occupying the 
same position in the HR diagram.
 Since these differences are larger that COROT's 
accuracy, one could use COROT to test these models.

\section{Conclusions \& future developments}

 The seismic characteristics of the PMS models evaluated 
here reflect their internal structure.  
 Therefore, as the models contract towards the ZAMS 
their structure and seismic properties change.
 In general terms, the evolution of $\Delta\nu$ and 
$r_i$ shows no signi\hbox{ficant} dependence on the 
model's mass and metallicity. 
 However these variations are correlated to some 
extent with the model's effective temperature. 
 Indeed, in the range of metal abundances \hbox{analy}sed 
here, the knowledge of  star's effective temperature 
(with a 100K uncertainty) allows us to infer its 
$\Delta\nu/$$<$$\rho$$>^{1/2}$ ratio with a 
precision between 2 and 5.5\%.

 Due to a degeneracy between the seismic properties 
($\delta\nu$ \& $r_i$) of some models, C-D diagrams 
are only useful for testing PMS models with less 
than 1.3M$_\odot$. This degeneracy is \hbox{broken by}
taking into account their temperature. 
 Consequently,~ $T_{eff}$ vs. $r_i$ 

\begin{table}[htbp]
\tabcolsep=1.mm
  \begin{center}  
    \caption{Models that occupy the same position in the 
HR diagram and have different seismic characteristics.}
 \begin{tabular}{ c c c c c c c } \hline \hline
M/M$_\odot$ & log(T$_{eff}$) & log(L/L$_\odot$) & Z & age(Myr) &
 $\Delta\nu$($\mu$Hz) & $r_i$(s$^{-1}$) \\ \hline 
2.5  &   3.75 &   1.16 &   0.01 &   1.66  &   27.9 &   -366.4 \\ \hline
2.5  &   3.75 &   1.16 &   0.02 &   2.46  &   28.6 &   -454.0 \\ \hline
\multicolumn{5}{r}{relative difference}   &  2.3\% &   19.3\% \\ \hline  
2.6  &   3.86 &   1.68 &   0.01 &   2.25  &   24.1 &   -762.4 \\ \hline
2.7  &   3.86 &   1.68 &   0.02 &   2.74  &   24.4 &   -799.1 \\ \hline
\multicolumn{5}{r}{relative difference}   &  1.3\% &    4.6\% \\ \hline 
2.6  &   4.05 &   1.92 &   0.01 &   2.82  &   53.3 &  -1919.7 \\ \hline
3.1  &   4.05 &   1.92 &   0.02 &   2.58  &   59.2 &  -1662.3 \\ \hline
\multicolumn{5}{r}{relative difference}   &  9.5\% &  13.4\%  \\ \hline
\end{tabular}
    \label{tab:1}
  \end{center}
\end{table}

\noindent diagrams can be used to analyse stars more massive 
than 1.3M$_\odot$. 
\noindent On the other hand, the accuracy in individual 
frequency determination that COROT can achieve is enough 
to test several PMS evolutionary models. 
 This result supports the study of young solar-type pulsators. 

 In the near future we will analyse the effect that 
stellar parameters, such as the mixing and the overshooting,  
have on the seismic properties of low/intermediate mass PMS 
stars. 
 The seismic properties expected for each model will be 
estimated u\hbox{sing} the  ADIPLS pulsation code 
(Christensen-Dalsgaard \cite{dalsgaard82}).


\begin{acknowledgements}
This work was supported by 
Funda\c{c}\~{a}o para a Ci\^{e}ncia e a Tecnologia 
and FEDER (through POCI2010) through 
project \hbox{POCI/CTE-AST/57610/2004}.
 I would also like to thank  M. Monteiro, J. Fernandes 
and the anonymous referee for their useful remarks and 
J. Marques for providing his models.
\end{acknowledgements}



\begin{thebibliography}{99}

\bibitem[2005]{antia05} Antia, H.~M., 2005, JApA, 26, 161 

\bibitem[2006]{antonello06} Antonello, E., 
  Mantegazza, L., Rainer, M., \& Miglio, A.\ 2006, A\&A, 445, L15 

\bibitem[2001]{baglin01} Baglin, A., Auvergne, M., Catala, C., 
  Michel, E., COROT Team, 2001, ESA-SP 464, 395

\bibitem[2003]{bedding03} Bedding, T.~R., Kjeldsen, H., 2003, PASA, 20, 203

\bibitem[1972]{breger72} Breger, M., 1972, ApJ, 171, 539 

\bibitem[2004]{browning04} Browning, M.~K., Brun, 
A.~S., \& Toomre, J.\ 2004, ApJ, 601, 512 

\bibitem[1982]{dalsgaard82} Christensen-Dalsgaard, J., 1982, MNRAS, 199, 735 

\bibitem[1983]{dalsgaard83} Christensen-Dalsgaard, J., Frandsen, S., 
  1983, SoPh, 82, 469 

\bibitem[1984]{dalsgaard84} Christensen-Dalsgaard, J., 1984, 
         Space Research in Stellar Activity and Variability, 11

\bibitem[1977]{goldreich77} Goldreich, P.,Keeley, D.~A., 1977, ApJ, 
  212, 243 

\bibitem[2002]{guenther02} Guenther, D.~B., 2002, ApJ, 569, 911 

\bibitem[1995]{kjeldsen95} Kjeldsen, H., Bedding, T.~R., 1995, A\&A, 293, 87 

\bibitem[1998]{marconi98} Marconi, M., Palla, F., 1998, ApJ, 507, L141 

\bibitem[2004]{marques04} Marques, J.~P., Fernandes, J.,
         Monteiro, M.~J.~P.~F.~G., 2004, A\&A, 422, 239

\bibitem[2006]{mazumdar06} Mazumdar, A., Basu, 
  S., Collier, B.~L., Demarque, P.\ 2006, MNRAS, 372, 949 

\bibitem[2002]{monteiro02} Monteiro, M.~J.~P.~F.~G., 
         Christensen-Dalsgaard, J., Thompson, M.~J, 2002, 
         ESA SP-485: Stellar Structure and Habitable Planet Finding, 
         291-298

\bibitem[1997]{morel97}  Morel, P., 1997, A\&AS, 124, 597

\bibitem[2003]{pinheiro03} Pinheiro, F.~J.~G., Folha, D.~F.~M., 
        Marconi, M., Ripepi, V., Palla, F., Monteiro, M.~J.~P.~F.~G.,  
        Bernabei, S., 2003, A\&A, 399, 271-274

\bibitem[2006]{pinheiro06} Pinheiro, F.~J.~G., 2006, 
       2005: Past Meets Present in Astronomy and Astrophysics, 23   

\bibitem[2004]{ripepi04} Ripepi, V., Marconi, M., 2004, 
  ESA SP-538: Stellar Structure and Habitable Planet Finding, 397 

\bibitem[2007]{ripepi07} Ripepi, V., Bernabei, S., Marconi, M., 
  Ruoppo, A., Palla, F., Monteiro, M.~J.~P.~F.~G., Marques, J.~P., 
  Ferrara, P., Marinoni, S., Terranegra, L., 2006, A\&A, 462, 1023

\bibitem[2003]{roxburgh03} Roxburgh, I.~W., Vorontsov, S.~V., 
         2003, A\&A, 411, 215

\bibitem[2001]{samadi01} Samadi, R., Goupil, M.-J., 2001, \aap, 370, 136 

\bibitem[2005]{samadi05} Samadi, R., Goupil, M.-J., Alecian, E., Baudin, 
    F., Georgobiani, D., Trampedach, R., Stein, R.,  Nordlund, {\AA}, 
    2005, JApA, 26, 171 

\bibitem[1980]{tassoul80} Tassoul, M., 1980, ApJS, 43, 469

\end{thebibliography}
\end{document}